\documentclass[final,5p,times,twocolumn]{elsarticle}

\usepackage{graphicx}
\usepackage[utf8]{inputenc}
\usepackage{units}
\usepackage{amssymb}
\usepackage{amsmath}

\usepackage{listings}
\usepackage{picture}
\usepackage{stfloats}
\usepackage{color}

\journal{Computer Physics Communications}

\begin{document}

\begin{frontmatter}



\title{Massively parallelized replica-exchange simulations of polymers on GPUs}


\author[juelich]{Jonathan Gross}
\author[leipzig]{Wolfhard Janke}
\author[juelich]{Michael Bachmann}

\address[juelich]{Soft Matter Systems Research Group, Institut f\"ur Festk\"orperphysik (IFF-2) and Institute for Advanced Simulation (IAS-2),\\ Forschungszentrum J\"ulich, D-52425 J\"ulich, Germany}
\address[leipzig]{Institut f\"ur Theoretische Physik and Centre for Theoretical Sciences (NTZ), Universit\"at Leipzig, Postfach 100920, D-04009 Leipzig, Germany}

\begin{abstract}
We discuss the advantages of parallelization by multithreading on graphics processing units (GPUs) for parallel tempering Monte Carlo computer simulations of an exemplified bead-spring model for homopolymers.
Since the sampling of a large ensemble of conformations is a prerequisite for the precise estimation of statistical quantities such as typical indicators for conformational transitions like the peak structure of the specific heat, the advantage of a strong increase in performance of Monte Carlo simulations cannot be overestimated.
Employing multithreading and utilizing the massive power of the large number of cores on GPUs, being available in modern but standard graphics cards, we find a rapid increase in efficiency when porting parts of the code from the central processing unit (CPU) to the GPU.
\end{abstract}


\end{frontmatter}


\section{Introduction}
\label{sec:intro}

Computer simulations have become a fundamental pillar in physics. 
In particular, computer simulations are frequently the only choice to understand physical properties of complex and cooperative behavior of systems which require a detailed modeling. 
This is certainly apparent in structural biophysics and polymer physics, where effective many-body interactions and disorder effects cannot be tackled by means of analytical approaches alone. 
Even for simplistic models, computation time can become interminable if a large amount of data is needed as, e.g., in statistical physics.

Despite large advances in the design of central processing unit (CPU) architectures most of the above-mentioned needs could not be met by simulations on single CPU systems. Several approaches to speed up simulations have been employed, e.g., parallel computing using \emph{message passing} on clusters or \emph{multithreaded programming} on multicore CPUs.

Graphics processing units (GPUs) have become very powerful, in recent years, driven by the professional computer gaming industry. 
GPUs possess a massively parallel architecture.
With the latest release of NVIDIA's convenient programming language CUDA, GPUs have become popular in scientific computing.
GPU computing finds its application in many fields, such as astronomy \cite{Ford2009406, Harris2008}, medicine \cite{0031-9155-54-20-017, 0031-9155-54-21-008}, time series analysis for financial markets \cite{0295-5075-82-6-68005}, molecular dynamics simulations \cite{Meel:2008, friedrichs2009}, Monte Carlo studies of spin systems \cite{PhysRevE.80.051117, weigeltbp, Preis20094468}, and Quantum Monte Carlo applications \cite{Meredith2009151}.
We are interested in the thermodynamical properties of polymer models, both on lattice \cite{prlBachmannJanke, ISI:000220456400052} and off-lattice \cite{PhysRevE.71.031906, Schnabel2009201}.
Previous studies \cite{Schnabel2009201, Schnabel2009, PhysRevE.81.011802} of an elastic polymer model revealed a complex, chain-length dependent structural transition behavior. 
For relatively high temperatures, the polymer chain has a wide spread coil-like structure. 
At the $\Theta$-point -- where monomer-monomer attraction and repulsion by volume exclusion is just balanced -- the polymer collapses from the random coil to more globular structures.
In this globular ``phase'', there is no internal structure.
This is comparable to a liquid. At even lower temperatures, sort of a freezing transition is observed.

The purpose of this paper is to show that GPU simulations can also quite efficiently be performed for off-lattice polymer models without any need of highly sophisticated tricks of implementation.
By employing a straightforward implementation, of massive parallelization provided by GPUs, we investigate the possible speed-up for replica-exchange Monte Carlo simulations of an off-lattice model for elastic polymers.

The paper is organized as follows.
Section~\ref{sec:cuda} describes details of the GPU architecture and CUDA.
In Section~\ref{sec:model} we give a brief introduction of the investigated model and the simulation technique used.
The results of our studies are presented in Section~\ref{sec:results}.
A summary of our findings in given in Section~\ref{sec:summary}.

\section{General-Purpose Computation on Graphics Processing Units}
\label{sec:cuda}
Since massively parallel general-purpose computation on GPUs is not standard, despite the large number of applications in the past few years, let us review the main features of multithreaded GPU architectures and the most frequently used specific language CUDA.

\subsection{GPU Architecture} 
\label{sub:cuda}

A GPU is composed of a number of streaming multiprocessors (SM) with on-chip shared memory only visible to that SM and a large global memory, often with sizes in the range of $1-4\unit{GB}$, in today's graphics card architectures.
The kernel -- the main function of a GPU program -- runs the same code in parallel on a number of threads given by the grid and block layout.

\begin{figure}[t]
  \centering
    \includegraphics[width=2.9in]{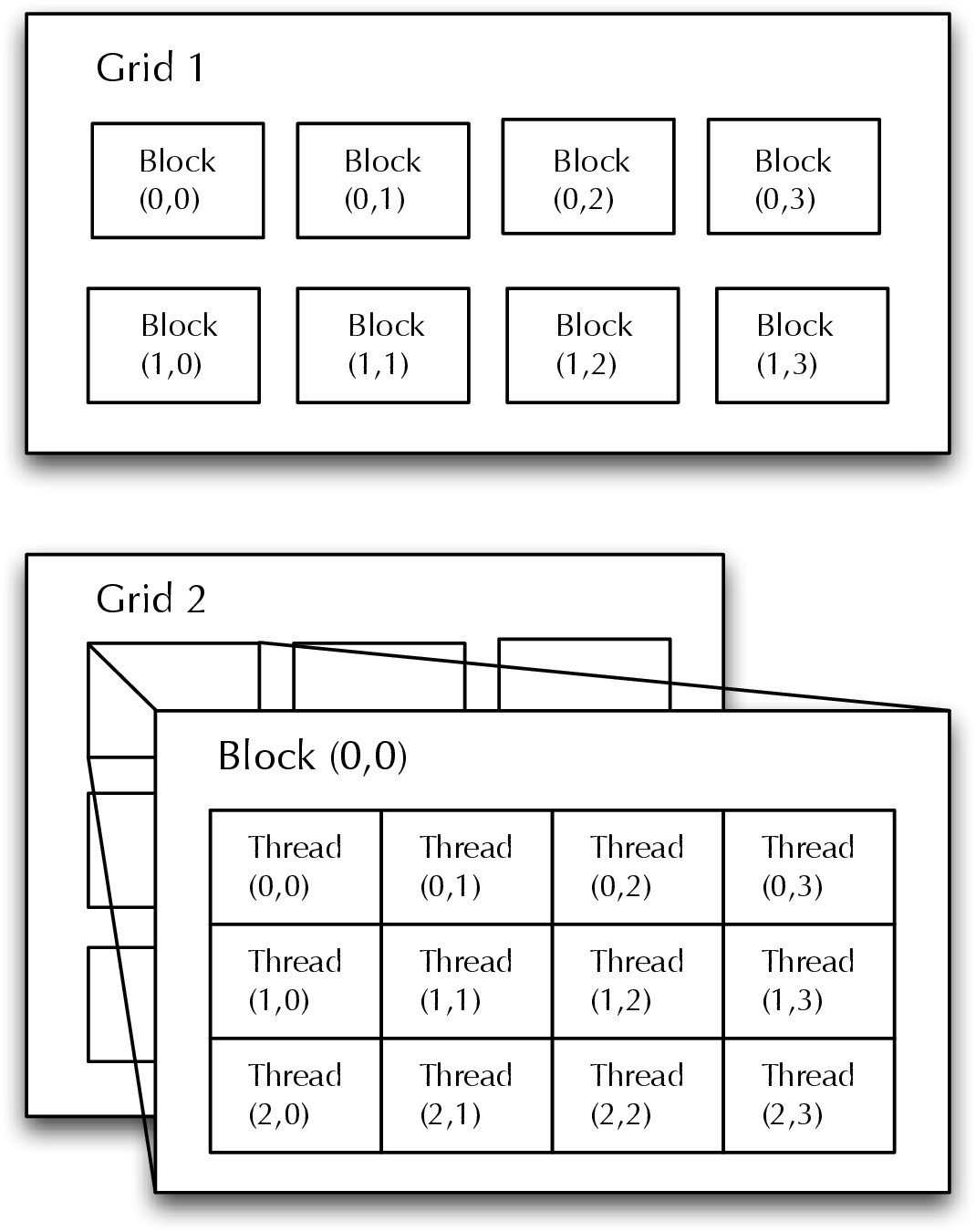}
  \caption{Grids with thread blocks.}
  \label{fig:pics_grid}
\end{figure}

The grid of independent thread blocks (Fig.~\ref{fig:pics_grid}) can be ordered in one or two dimensions.
It is possible to launch the same kernel or others with a differing grid layout within the same program successively, with the Fermi cards even concurrently.
Threads, i.e., individual processes, can be arranged within these blocks in up to three dimensions with a maximum size of $512\times512\times64$ on NVIDIA's GT200-based cards or $1024\times1024\times64$ on Fermi-based cards.
However, it is also limited to $512$ or $1024$ threads per block, respectively. Each thread can be identified by a unique id, which can be calculated from its coordinates within the grid.
At execution time, $32$ threads are grouped together in a \emph{warp}.
Those warps are then assigned to a SM.
For maximum performance, it is best to start many more threads than cores exist on the device.
The same program can run on a wide range of CUDA-capable devices with different number of cores, because the distribution of warps to cores is done on the device.
This feature is called \emph{transparent scalability} \cite{kirk2010cuda}.
Also, if certain threads within a warp are reading data from global memory, the device is able to hide the memory latency by executing a warp of threads that does not have to wait for another operation.

\begin{figure}[t]
  \centering
    \includegraphics[width=0.9\columnwidth]{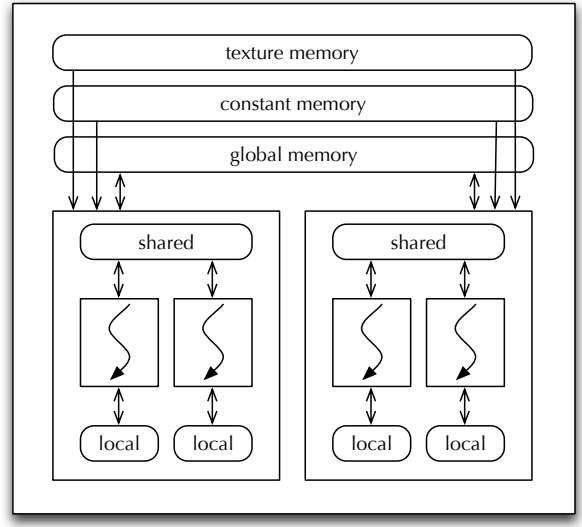}
  \caption{Memory layout on a GPU device.}
  \label{fig:pics_memory}
\end{figure}

Graphics cards come with several types of memory (Fig.~\ref{fig:pics_memory}) accessible to the program.
The largest is the \emph{global memory}, which can be as large as $4\unit{GB}$ in professional special purpose cards like the Fermi series from NVIDIA.
The global memory can be read and written by specific functions from the CPU -- also called \emph{host} --, and every thread on the GPU -- also called \emph{device} -- has read and write access to global memory.
This large memory is necessary since it is not possible for the device to access the RAM of the host.
This means all data that needs to be processed by the GPU has to be copied to the device for calculation and copied back to the host for evaluation.
A downside of the global memory is its high latency. 

The \emph{constant memory} supports short latency read-only access by the device if all threads read from the same location in memory.
For specific data types, the \emph{texture memory} is available.
\emph{Registers} and \emph{shared memory} are fast on-chip memories.
Access to these types of memory is usually a lot faster than global memory.
Registers are assigned per thread and only accessible by that thread.
Shared memory is allocated to a thread block within a SM and all threads in this block can read and write to that memory.
The number of available registers is limited to $16384$ on GT200-based chips and is twice as large on new Fermi-based cards.
The size of the shared memory was increased from $16\unit{KB}$ on GT200 to $48\unit{KB}$ on Fermi cards \cite{fermi}.

\subsection{Compute Unified Device Architecture: CUDA}
With the release of NVIDIA's set of programming tools (CUDA) in 2007, the exploitation of the potential of GPUs has become more feasible.
The CUDA toolkit comes in two variants, a high-level C/C++ interface to GPU functions, the so-called CUDA runtime API and a more low-level programming layer, the CUDA driver API, which is closer to hardware.
The toolkit contains a set of extensions to the C programming language to accomplish the most common tasks in GPU programming, like memory management and operations, but also new data types for mathematical calculations.
For a detailed description of the CUDA programming language, see~\cite{CUDAGuide3.0}.
The drawback is that CUDA currently only runs on NVIDIA hardware.
However, there is a vendor-independent alternative in active development, called OpenCL (Open Computing Language)\footnote{http://www.khronos.org/opencl/}.
Its programming interface is comparable to the driver API of NVIDIA's CUDA. 

\begin{figure}[t]
  \centering
    \includegraphics[width=\columnwidth]{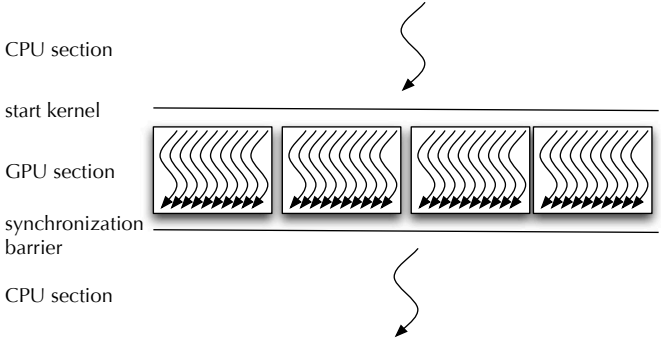}
  \caption{Sequence of a parallel GPU program.}
  \label{fig:paper_pics_threads}
\end{figure}

Since at this time CUDA seems more mature than OpenCL and programming the CUDA runtime API is much more convenient than low-level programming, this is what we chose for our implementation.

A scheme of the sequence of a CUDA parallel program is shown in Fig.~\ref{fig:paper_pics_threads}. 
The program starts on the CPU like any other program with the initialization of variables and data.
With CUDA one also has to allocate and initialize all memories that are needed for the calculations on the GPU.
When all data is copied to the device, the kernel is started, and the program now spreads into parallel threads running on the GPU.
The execution of the kernel on the GPU is completely independent from the calling program, which could proceed in its own execution.
For this reason a synchronization barrier has to be implemented in the main program to wait for the calculations on the device to finish.
After the kernel finishes its execution, the results of the computations are copied back to the host memory for further processing.

\subsection{Random Number Generation} 
\label{sub:random_numbers}
Monte Carlo simulations require a multitude of random numbers.
In our implementation, the required amount of random numbers for a single kernel call is generated on the CPU and copied to global memory before the actual execution of the kernel.
Each thread block then copies the needed numbers to its shared memory during the execution.
This is not the most optimal solution, because the memory transactions between global and shared memory slow down the overall execution time.
It would be more efficient to run an independent random number generator on each thread block that fits entirely into shared memory.
With the advancing size of the shared memory on the device, this is an option for future implementations.

\section{GPU Simulations of an Elastic Flexible Polymer Model}
\label{sec:model}

\subsection{Polymer Model} 
\label{sub:polymer_model}

As an example for a molecular system, we consider an elastic, flexible bead-spring homopolymer chain.
All monomers interact via a shifted and truncated pairwise Lennard-Jones potential:
\begin{equation}
  E_{\textrm{\tiny{LJ}}}^{\textrm{\tiny{mod}}}(r_{ij})=E_{\textrm{\tiny{LJ}}}(\min(r_{ij}, r_c)) - E_{\textrm{\tiny{LJ}}}(r_c),
\end{equation}
\begin{equation}
  E_{\textrm{\tiny{LJ}}}(r_{ij})=4\epsilon\left[\left(\frac{\sigma}{r_{ij}}\right)^{12}-\left(\frac{\sigma}{r_{ij}}\right)^6\right],
\end{equation}
where $r_{ij}$ is the distance between two monomers $i$ and $j$.
The Lennard-Jones parameters are $\epsilon=1$ and $\sigma=2^{-1/6}r_0$, the minimum of the potential is at $r_0=0.7$ and the cutoff radius is $r_c=2.5\sigma$.
To model the bonds between adjacent monomers, we use the finitely extensible nonlinear elastic (FENE) anharmonic potential
\begin{equation}
  E_{\textrm{\tiny{FENE}}}(r_{ii+1})=-\frac{K}{2}R^2\log\left[1-\left(\frac{r_{ii+1}-r_0}{R}\right)^2\right],
\end{equation}
which has its minimum also at $r_0$ and diverges for $r \rightarrow r_0 \pm R$ with $R=0.3$.
The spring constant $K$ for the FENE bonds equals $40$ \cite{Schnabel2009201, Schnabel2009}.

The total energy of a conformation $\mathcal{C}=(\mathbf{r}_1,\dots,\mathbf{r}_N)$ is then given by 
\begin{equation}
  E(\mathcal{C})=\frac{1}{2}\sum_{\substack{i,j=1\\i \ne j}}^N E_{\textrm{\tiny{LJ}}}^{\textrm{\tiny{mod}}}(r_{ij}) + \sum_i^{N-1} E_{\textrm{\tiny{FENE}}}(r_{ii+1}).
\end{equation}
The conformational behavior of elastic polymers in this model has already been investigated in detail in Refs. \cite{Schnabel2009201, Schnabel2009}. 

\subsection{Monte Carlo Update}
The local update used throughout the simulation is the following:
A random monomer of the chain is picked and a shift of its coordinates by a random vector is proposed.
The components of this displacement vector are uniformly distributed random numbers within the interval $[-0.01,0.01]$.

For a given inverse thermal energy $\beta=1/k_BT$ at temperature $T$, this proposal is accepted or rejected according to the standard Metropolis \cite{metropolis1953} criterion
\begin{equation}
  p=\min\left(1,\exp\left[-\beta(E_{\textrm{\tiny{new}}}-E_{\textrm{\tiny{old}}})\right]\right),
\end{equation}
where $E_{\textrm{\tiny{new}}}$ is the energy of the proposed new conformation and $E_{\textrm{\tiny{old}}}$ the energy of the original one.

\subsection{Replica-exchange Method}
In the replica-exchange parallel tempering \cite{PhysRevLett.57.2607, Geyer1991, partemp} method, a simulation of $n_r$ copies, i.e., replicas of the same system, is run at different temperatures.
After a certain number of Monte Carlo updates an attempt to exchange the conformations of neighboring replicas $i$ and $i+1$ is performed.
The probability to accept such an exchange is given by
\begin{equation}
  \label{eq:ptswap}
  p=\min\left(1,\exp\left[(E_i-E_{i+1})(\beta_i-\beta_{i+1})\right]\right).
\end{equation}
This heats up and cools down every copy of the system, which helps to avoid barriers in the free-energy landscape and to reduce autocorrelation times.
Thus in principle, the effective statistics can be increased.

In this study we show that parallel tempering can quite efficiently be run on massively parallelized architectures, such as GPUs.

\subsection{CUDA Implementation}
\label{sec:method}
In this section, we will focus the attention on implementation details.
The following listings contain CUDA specific syntax emphasizing the most relevant parts of a CUDA program.
All calculations are performed using single-precision floating-point operations, because support for double-precision is not generally available.
Double-precision performance is to be further improved in future chip generations.
Listing \ref{kernelcall} shows how the main function of a GPU program is invoked.
First, one needs to set up the dimensions for the grid of threads.
In this case, the dimension of the grid (\textsf{dimGrid}) is set to the number of replicas.
This means that every replica of the systems runs independently in its own thread block.
The size of such a thread block \textsf{dimBlock} is set to a constant \textsf{BS} which depends on the number of beads in the polymer.
Details will be discussed in Section~\ref{sub:performance_comparison}.
\lstset{language=C,basicstyle=\sffamily\scriptsize,showspaces=false,showstringspaces=false,showtabs=false,frame=single,tabsize=2,captionpos=b,breaklines=true,breakatwhitespace=false,escapeinside={\%*}{*)},
caption=Specification of the thread layout and kernel call within the main program.,label=kernelcall}
\begin{lstlisting}
dim3 dimGrid(NCONFS);
dim3 dimBlock(BS);
run<<<dimGrid,dimBlock>>>
   (d_confs, d_rnds, d_energies, d_rees, d_rgyrs);
cudaThreadSynchronize();
\end{lstlisting}
After setting up the grid layout, the kernel \textsf{run} is called with the given layout embraced by triple chevrons and a list of arguments.
The same code is then executed by each thread.

Because a kernel function call returns immediately, it is necessary to call \textsf{cudaThreadSynchronize()} as a synchronization barrier.
Thus the CPU waits for the GPU to finish the calculations before proceeding.
In our implementation the exchange of replicas is done on the CPU, while the GPU is used for the Metropolis algorithm with the expensive energy calculation.  
A pruned record of the insides of the kernel is shown in Listing~\ref{kernel}. 
\lstset{language=C,numbers=left,numberstyle=\tiny,stepnumber=2,numbersep=5pt,basicstyle=\sffamily\scriptsize,showspaces=false,showstringspaces=false,showtabs=false,frame=single,tabsize=2,captionpos=b,breaklines=true,breakatwhitespace=false,escapeinside={\%*}{*)},
caption=The kernel function -- showing the usage of shared memory and the main work loop. \textsf{TX} is the unique id for each thread.,label=kernel}
\begin{lstlisting}
__global__ void run
  (Polymer* d_confs, float* d_rnds, float* d_energies, 
   float* d_rees, float* d_rgyrs) 
{
  int id = blockIdx.x;
  __shared__ Polymer ps;
//.. initialization of some variables
  if (TX == 0) {
    ps = d_confs[id];
  }
  __syncthreads();
  
  while (n < NSWEEPS) {
    oneSweep(&ps, rnds, n);
    ener = energy(&ps);
    if (TX == 0) {
      Ree = endToEndDistance(&ps);
      Rgyr = radiusOfGyration(&ps); 
      d_energies[n+offset] = ener;
      d_rees[n+offset] = Ree;
      d_rgyrs[n+offset] = Rgyr;
      n++;
    }
    __syncthreads();
  }
  if (TX == 0) {
    d_confs[id] = ps;
  }
  __syncthreads();
}
\end{lstlisting}
In line 5 of Listing~\ref{kernel}, the index of the current thread block is assigned to an integer variable and is used to link this thread block to a specific replica.
The shared memory for a local copy of the replica is allocated in line 6.
This means that all threads within this block, and only those, have fast access to the copy.
The actual copy process is shown in lines 6 -- 11, where only one thread is assigned to copy the polymer from global to shared memory.
The barrier in line 11 lets all other threads of the block wait for the copying to finish, before proceeding with the actual calculation.
In line 15, the energy calculation is invoked.
The details of the parallel implementation and CUDA specifics are explained in Listings~\ref{FENE} and~\ref{lj}.
Again, only one thread is used in lines 16 -- 23 to collect statistics.
The local copy of the polymer is copied back to global memory in line 27 for further evaluation on the CPU.

Due to the fact that every thread executes the same code, it is possible to insert conditions based on the id of the thread to alter what each thread actually calculates.

Since in this model there are only pairwise interactions between monomers, it is possible to parallelize the calculation of the energy.
For the FENE part of the potential this is particularly straightforward, because only next neighbors are involved, see Listing~\ref{FENE}.
\lstset{language=C,basicstyle=\sffamily\scriptsize,showspaces=false,showstringspaces=false,showtabs=false,frame=single,tabsize=2,captionpos=b,breaklines=true,breakatwhitespace=false,escapeinside={\%*}{*)},
caption=Calculation of pairwise FENE interactions.,label=FENE}
\begin{lstlisting}
if (TX < N-1) {
  r = distance(p, TX, TX+1);
  energy[TX]+=-0.5*K*R*R*__logf(1-((r-r0)/R)*((r-r0)/R));
}
__syncthreads();
\end{lstlisting}
For the Lennard-Jones part of the potential, it is not that trivial to get the indices of all possible neighbors.
The calculation of the relevant indices is included in Listing~\ref{lj}.
\lstset{language=C,basicstyle=\sffamily\scriptsize,showspaces=false,showstringspaces=false,showtabs=false,frame=single,tabsize=2,captionpos=b,breaklines=false,breakatwhitespace=false,escapeinside={\%*}{*)},
caption=Calculation of pairwise Lennard-Jones potential.,label=lj}
\begin{lstlisting}
for (int i=0; i<N/2; i++) {
  if (TX < N) {
    if (TX > i) {
      index1 = i;
      index2 = TX;
    }
    else {
      index1 = N-2-i;
      index2 = index1+1+TX;
    }
    r = distance(p, index1, index2);
    if (r<rc && (i != N-2-i)) {
      float rs6 = __powf(sigma/r, 6.0f);
      energy[TX] += 4*epsilon*((rs6*(rs6-1))-E_lj_rc);
    }
    ...
__syncthreads();
\end{lstlisting}
For both parts of the potential, \emph{one pair} of monomers is assigned to \emph{one thread} to perform the actual calculation of the energy.
The results of those calculations are stored in the array \textsf{energy}.
When all threads are finished with their part a \emph{parallel reduction} is performed on this array to obtain the total energy.
Instead of using a single thread and a loop for the summation, multiple threads calculate different parts of the sum.
See \cite{reduction} for details.

\section{Results}
\label{sec:results}
\begin{table*}[t]
  \centering
  \caption{Specifications of the used hardware.}
  \begin{tabular}{l r r r r r r}
  \hline
  & reference CPU & GPU1 & GPU2 & GPU3 \\
  \hline \hline \\
  name & Xeon E5620 & Tesla C1060 & GTX285 & GTX480 \\
  \# processors & 1 & 30 & 30 & 15 \\
  \# cores per processor & 4 (only 1 used) & 8 & 8 & 32 \\
  RAM & 16384MB & 4096MB & 1024MB & 1536MB \\
  clock speed & 2.4GHz & 1.3GHz & 1.48GHz & 1.4GHz \\
  max. threads per block & - & 512 & 512 & 1024 \\
  shared memory size & - & 16kB & 16KB & 48kB \\
  registers per block & - & 16384 & 16384 & 32768 \\
  \hline
  \end{tabular}
  \label{table:hardware}
\end{table*}
\subsection{Thermodynamics}
\label{sub:thermodynamics}
Before discussing details of the efficiency of the GPU simulations, let us first briefly review the thermodynamical properties of elastic polymers with $13$ and $55$ monomers, which non-trivially freeze into icosahedral structures at low temperatures.

Figure~\ref{fig:pics_polymers_wkap} shows the specific heat of the polymers, given by
\begin{equation}
  \frac{C_V}{N}=\frac{1}{N}\frac{\partial\langle E \rangle}{\partial T}=\frac{\beta^2}{N}(\langle E^2\rangle-\langle E\rangle^2).
\end{equation}
There is a change in the monotonic behavior of the curve at approximately $T=1.0$ for the 13mer and approximately $T=1.5$ for the 55mer.
This is an indicator for a structural change within the polymer chain, in this case the $\Theta$-collapse, which describes the finite-system analog of the phase transition from random-coil conformations to globular shapes.

The very distinct peak at low temperatures, $T \approx 0.33$ for the 13mer and $T \approx 0.32$ for the 55mer, is a sign for the freezing transition. Below this temperature, the crystalline polymer has an icosahedral structure. These results conform with previous studies~\cite{Schnabel2009201, Schnabel2009}.

\setlength{\unitlength}{1in}
\begin{figure}[b]
  \centering
  \begin{picture}(\columnwidth,2.3)
    \put(0,0){\includegraphics[width=\columnwidth]{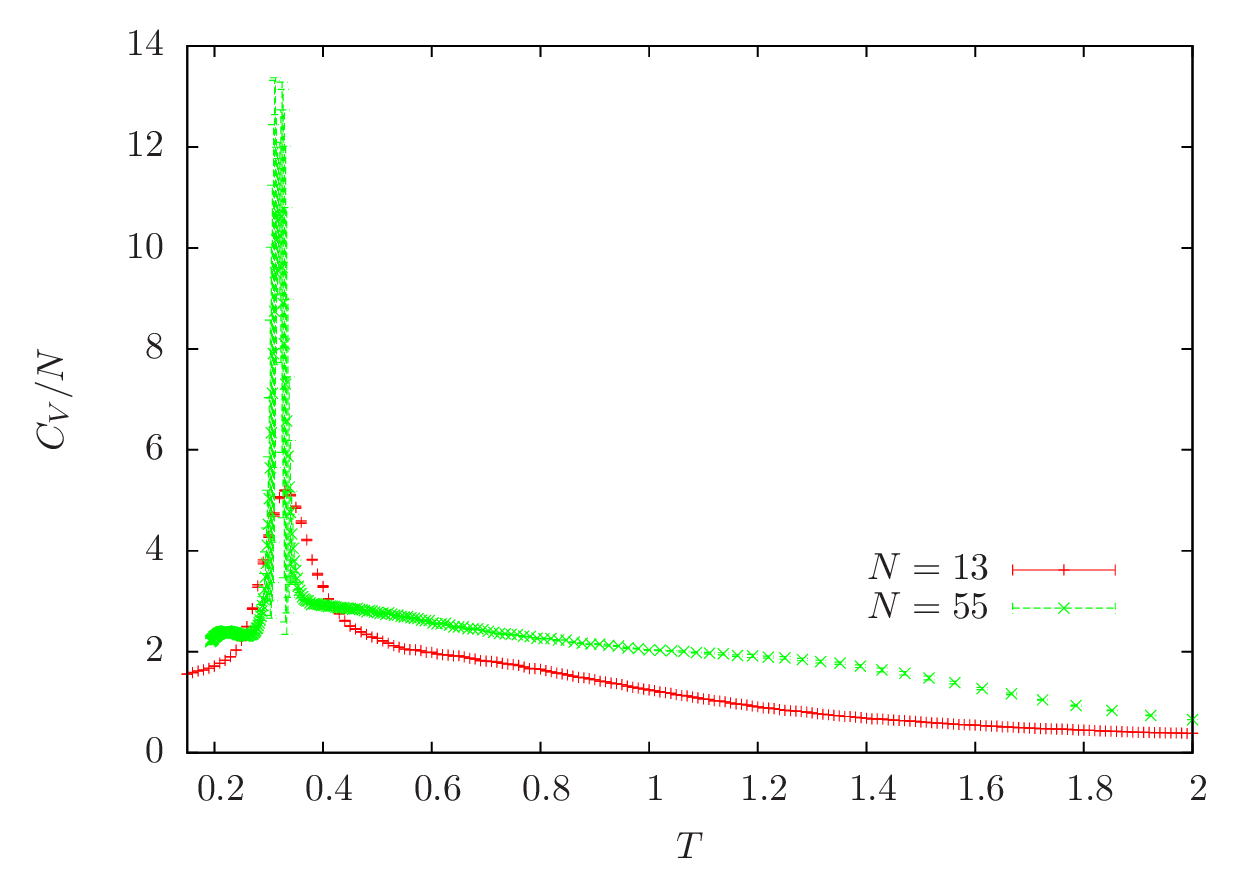}}
    \put(1.5,1.1){\includegraphics[width=1in]{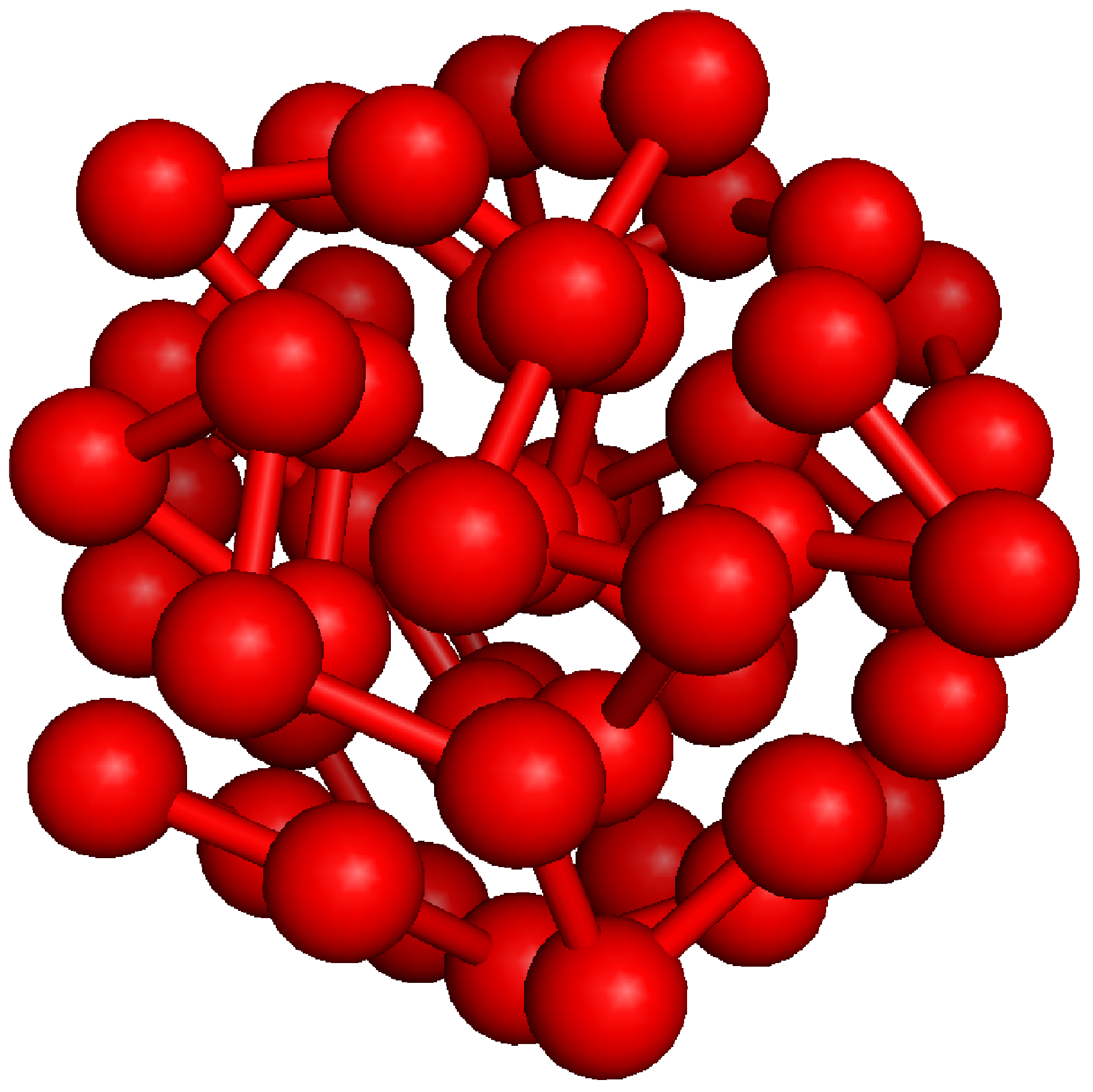}}
  \end{picture}
  \caption{Specific heat of an elastic polymer chain with lengths 13 and 55. The inset shows the icosahedral structure of the 55mer.}
  \label{fig:pics_polymers_wkap}
\end{figure}

\begin{figure}[b]
  \centering
    \includegraphics[width=\columnwidth]{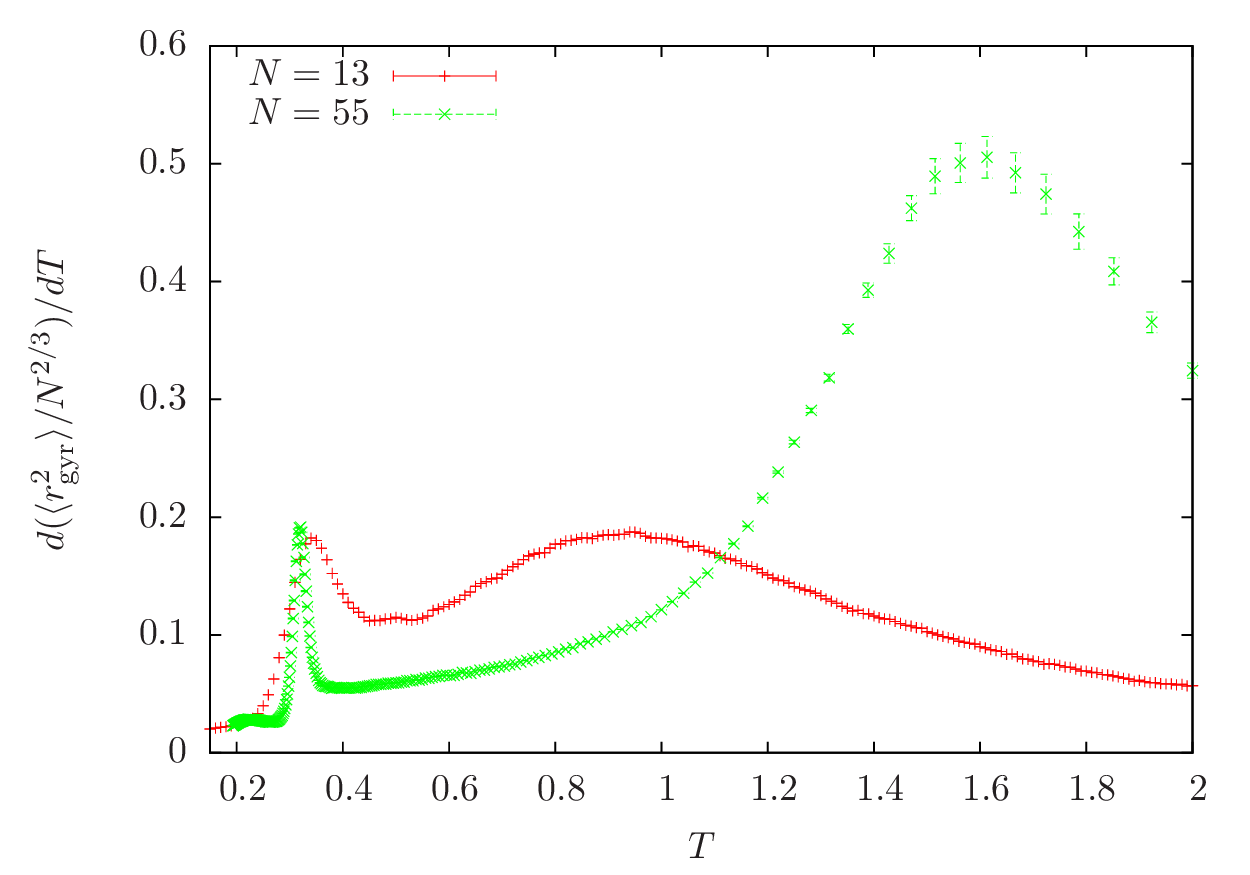}
  \caption{Fluctuation of the squared radius of gyration for chain lengths 13 and 55.}
  \label{fig:pics_polymers_rgyr}
\end{figure}

Figure~\ref{fig:pics_polymers_rgyr} shows the fluctuation of the squared radius of gyration, given by
\begin{equation}
   r_{\scriptsize\textrm{gyr}}^2 = \frac{1}{N}\sum_{k=1}^N(\vec{r}_k-\vec{r}_{\scriptsize\textrm{mean}})^2,
\end{equation}
where $\vec{r}_{\scriptsize\textrm{mean}}$ is the center of mass of the polymer.
The radius of gyration describes the mean distance of every monomer to the center of the polymer.
From its fluctuations, structural transitions can be identified. 
The peaks for the freezing transition approximately coincide with the peaks in the specific heat.
Much more distinct are the peaks that indicate the $\Theta$-collapse at higher temperatures, compared to the corresponding weak signals in the specific heat.

\subsection{Performance comparison}
\label{sub:performance_comparison}
To compare the performance of each implementation, we define a speed-up factor as follows,
\begin{equation}
  S_p = \frac{t_{\tiny\textrm{CPU}}}{t_{\tiny\textrm{GPU}}},
\end{equation}
where $t_{\tiny\textrm{CPU}}$ is the execution time on a single CPU core and $t_{\tiny\textrm{GPU}}$ is the runtime on the GPU.
All runtimes were measured with \textsf{cutil}-timers, which are wrapper functions to the standard C library call \textsf{gettimeofday}.
This ensures a consistent time measurement on a variety of systems, and to measure CPU and GPU time alike.
On the CPU side, only the time taken by the actual calculation was measured.
No initialization of variables or any file operations were included in the time measurement.
To consider the extra overhead which comes with GPU computing, the time taken to copy data hence and forth the device has been included along with the time taken for the sweeps.
For the speed comparison, only short runs of $4000$ sweeps per replica were performed.
Every $1000$ sweeps the replicas were copied back to the host to propose a swap of temperatures between neighboring copies according to Eq.~(\ref{eq:ptswap}).
The simulated system was of size $N=55$.
The specifications of the three tested GPUs are listed in Table~\ref{table:hardware}.
\begin{figure}[t]
  \centering
    \includegraphics[width=\columnwidth]{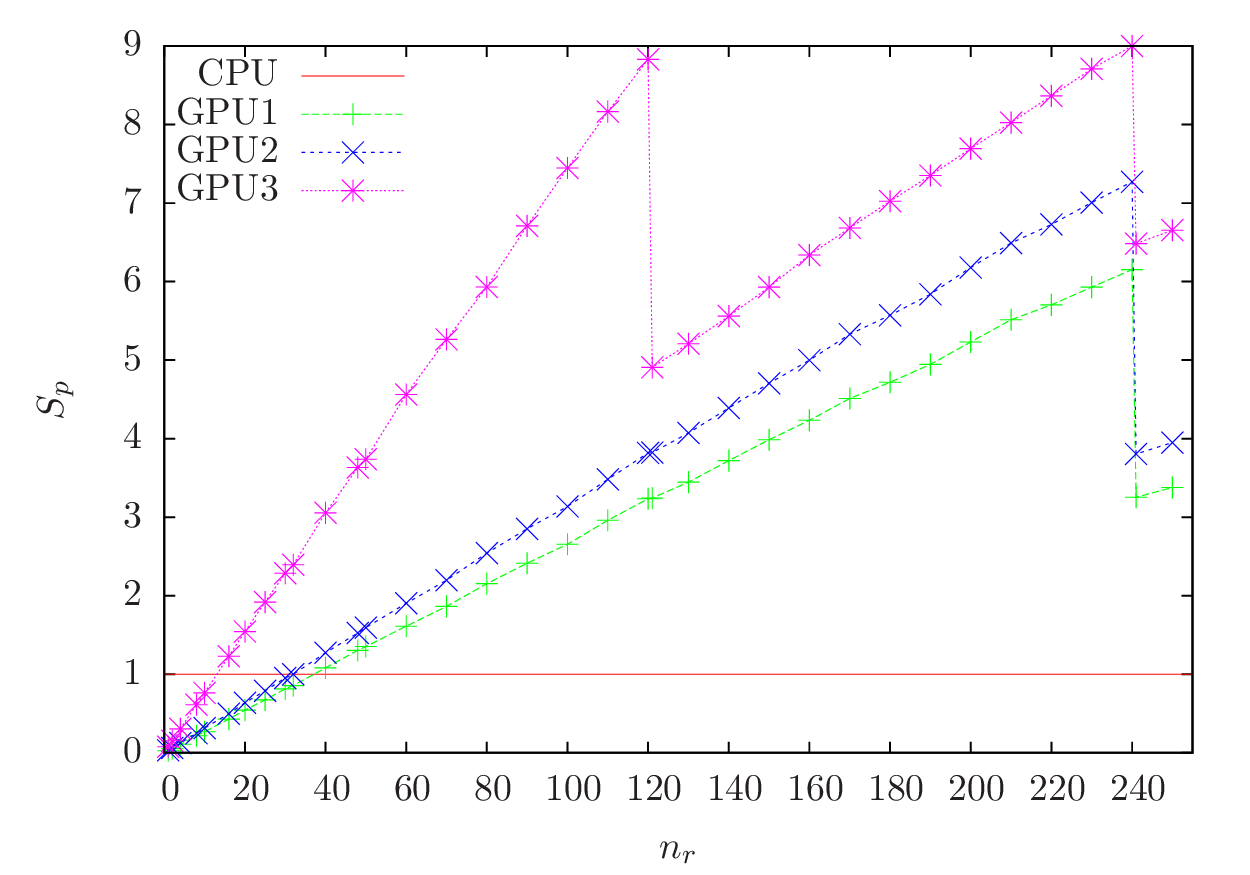}
  \caption{Speed-up factor $S_p$ vs. number of replicas $n_r$ for naive porting of CPU code to the GPU. Lines are only guides to the eye.}
  \label{fig:speedup_serial}
\end{figure}
First, naive portings of CPU to GPU codes already show maximum speed-ups around 6--7 for the GT200-based cards GPU1 and GPU2 compared to the reference implementation on the CPU.
With GPU3, based on the Fermi architecture, the maximum speed-up is about 9.
Figure~\ref{fig:speedup_serial} shows the dependency of the speed-ups on the number of replicas for different GPUs.
In the naive approach, each replica of the system was assigned to one thread block containing only one thread.  
Because of the embarrassingly parallel nature of the parallel tempering algorithm, it is possible to outperform a single CPU when more than 36 replicas of a system are simulated on GPU1, 32 replicas or 13 replicas for GPU2 or GPU3 respectively.
This is possible due to the large number of cores available on GPUs, even though their clock speeds are lower than that of modern CPUs. 

As mentioned in Section~\ref{sec:cuda}, the size of thread blocks is limited and threads on the GPU are bundled to groups called warps.
The maximum number of simultaneously active threads in a multiprocessor is 1024 (or 32 warps) for GT200-based cards (GPU1 and GPU2) and 1536 threads (or 48 warps) for Fermi-based GPU3.
These warps do not necessarily have to belong to the same thread block.
Thus also 16 warps from 2 different thread blocks can be active simultaneously in a single SM, or 3 blocks of 10 warps, or 4 blocks of 8 warps, and so on, up to 8 blocks of 4 warps.
This grouping of warps from up to 8 different thread blocks is a current limitation of the GPU architecture. 
The SM is not able to gather as many warps from different thread blocks until its warp or thread limit is reached; explaining the peaks in Figure~\ref{fig:speedup_serial}.
\begin{figure}[t]
  \centering
  \includegraphics[width=\columnwidth]{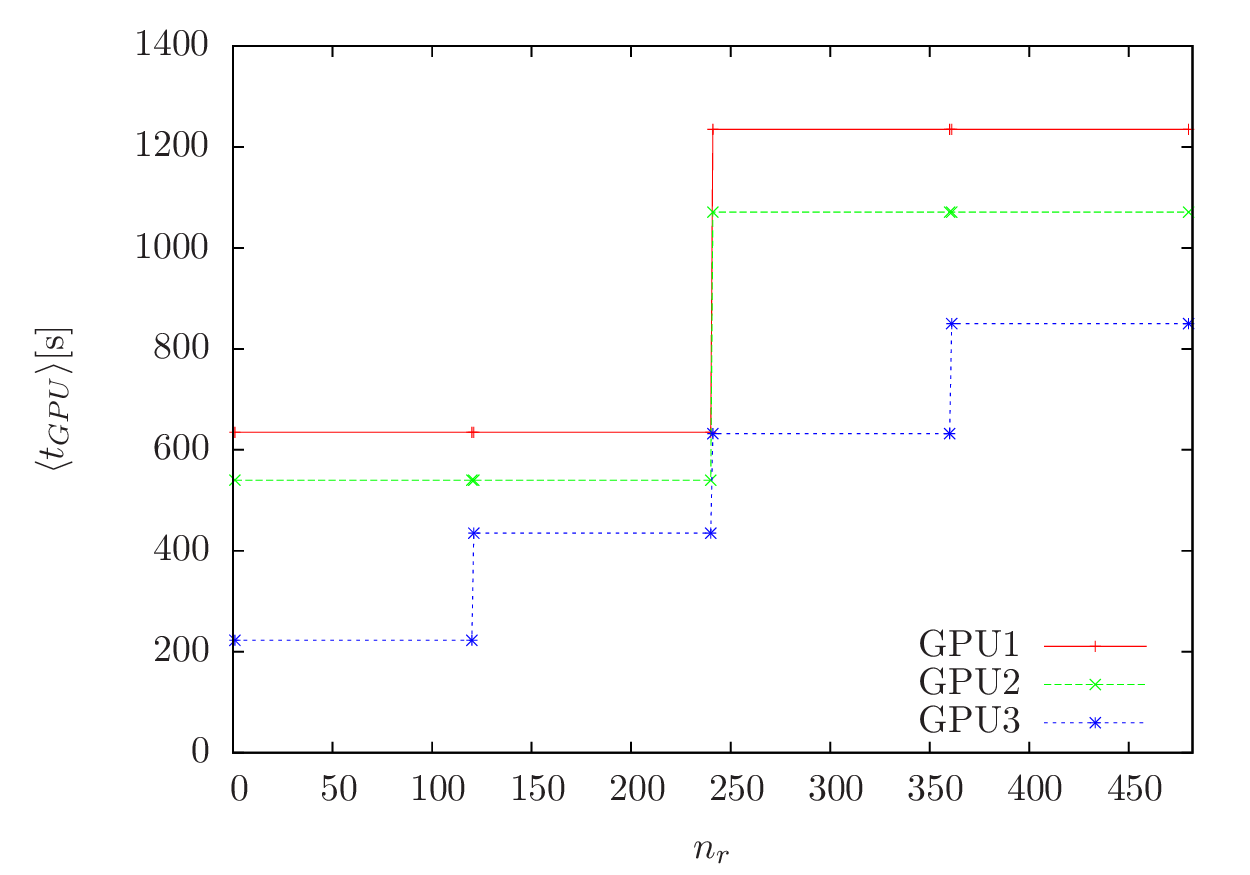}
  \caption{Average GPU time $\langle t_{GPU}\rangle$ in seconds vs. number of replicas $n_r$ for the naive GPU version.}
  \label{fig:runtimes1}
\end{figure}
Since there are 30 multiprocessors on GPU1 and GPU2, the maximum number of active warps of the device is reached for 240 thread blocks of 1 thread.
Each SM calculates the single threads of 8 different thread blocks.
GPU3 however has 15 multiprocessors, thus it is only capable of executing 120 blocks at a time with 1 thread per block.
The maximum speed-up for this thread layout is to be expected at multiples of 240 for GPU1 and GPU2, and multiples of 120 for GPU1.
Essentially thread blocks with threads other than multiples of 32 -- the warp size -- are not recommended at all.
Another interesting finding is that the total runtime of the kernel is the same, whether only one multiprocessor is busy and the others are idling, or all multiprocessors are equally busy.
This leads to a step-like graph, see Fig.~\ref{fig:runtimes1}, when plotting the kernel runtime versus the number of replica, i.e., the number of thread blocks.

With this in mind, an improved version was implemented with a parallel calculation of the energy function, as shown in Listings~\ref{FENE} and~\ref{lj}.
The thread block size in this version was set to 64 -- a multiple of the warp size, for better scheduling -- each block holding one replica.
This improved version also exploits low-latency memory access by using shared memory for storing the coordinates of the monomers.
Thus all threads within a block have fast access to them when needed for the calculation of their portion of the energy.
Performance is also gained by substituting calls to standard C math library with optimized CUDA versions \cite{CUDAGuide3.0}, see Listing~\ref{FENE} line 3 and Listing~\ref{lj} line 13.
Those usually have a lower precision than their counter-parts, but are executed faster.
\begin{figure}[t]
  \centering
  \includegraphics[width=\columnwidth]{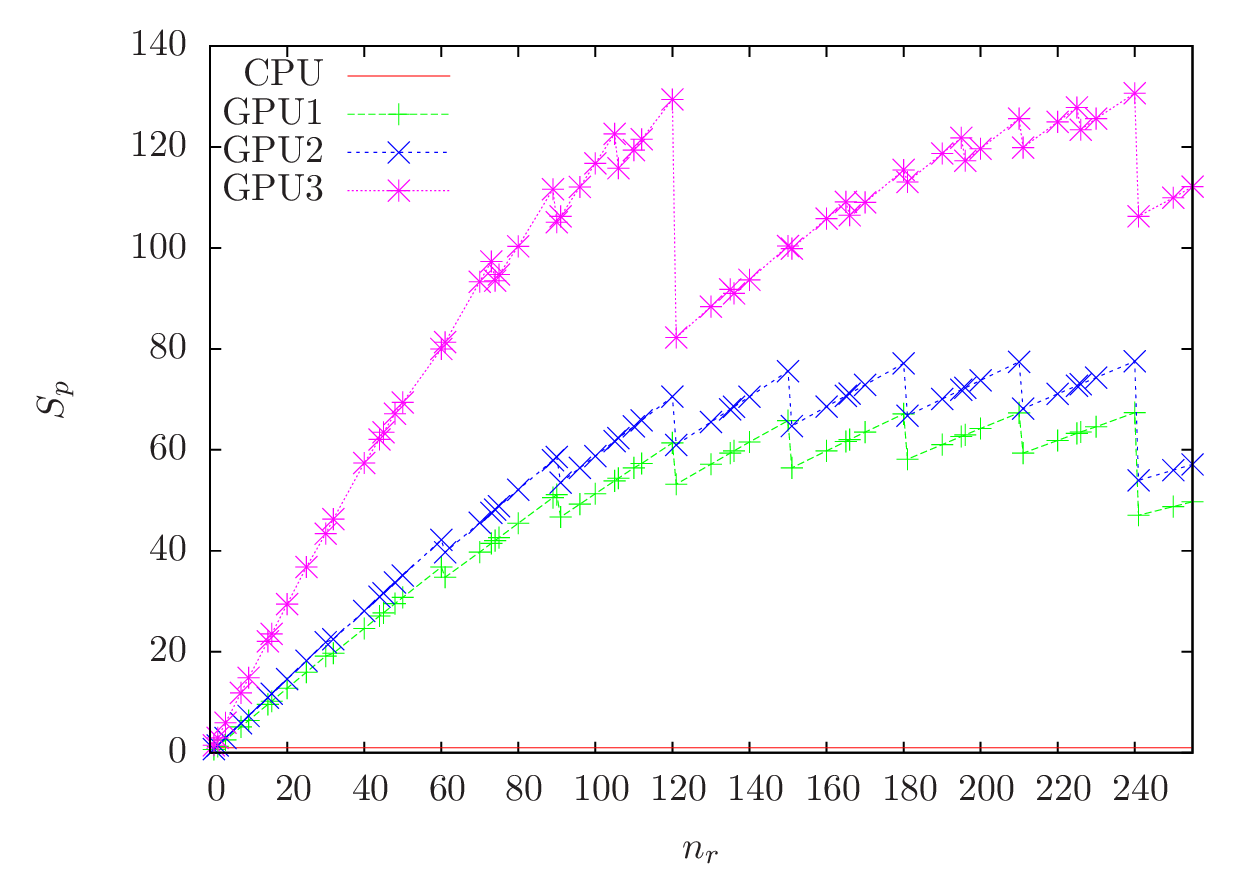}
  \caption{Speed-up factor $S_p$ vs. number of replicas $n_r$ for the GPU version with parallelized energy calculation. Drawn lines are only guides to the eye.}
  \label{fig:speedup_parallel}
\end{figure}
As shown in Fig.~\ref{fig:speedup_parallel}, this implementation is much faster than the CPU version, when more than 2 replica are simulated.
The maximum speed-up factor for GPU1 is 68, for GPU2 it is 78 and for GPU3 even 130. 
Again, for the two GT200-based cards, multiples of 240 threads blocks are a limit for the maximum speed-up.
With 240 active thread blocks of 64 threads each, there are 15360 threads running on the GPU.
The total number of threads divided by the number of SMs in these cards equals 512.
These 512 threads are a collection of 2 warps from 8 different thread blocks.
So, with this parametrization, the occupancy of the multiprocessors is only at $50\%$, since 1024 active threads per SM are possible with GT200-cards.
For GPU3, the first maximum is at 120 thread blocks, which complies with a total number of 7680 threads.
Even though GPU3 is capable of running 1536 threads in each of its 15 multiprocessors at a time, only 512 threads are active, due to the 8 thread block limitation (this equals an occupancy of $33\%$).
These occupancy values come from the fact that only 64 threads per replica are used for the energy calculation.
A different implementation with more threads or bigger systems sizes, which require more threads, could increase the occupancy of the multiprocessors and thus increase the efficiency.
\begin{figure}[b]
  \centering
  \includegraphics[width=\columnwidth]{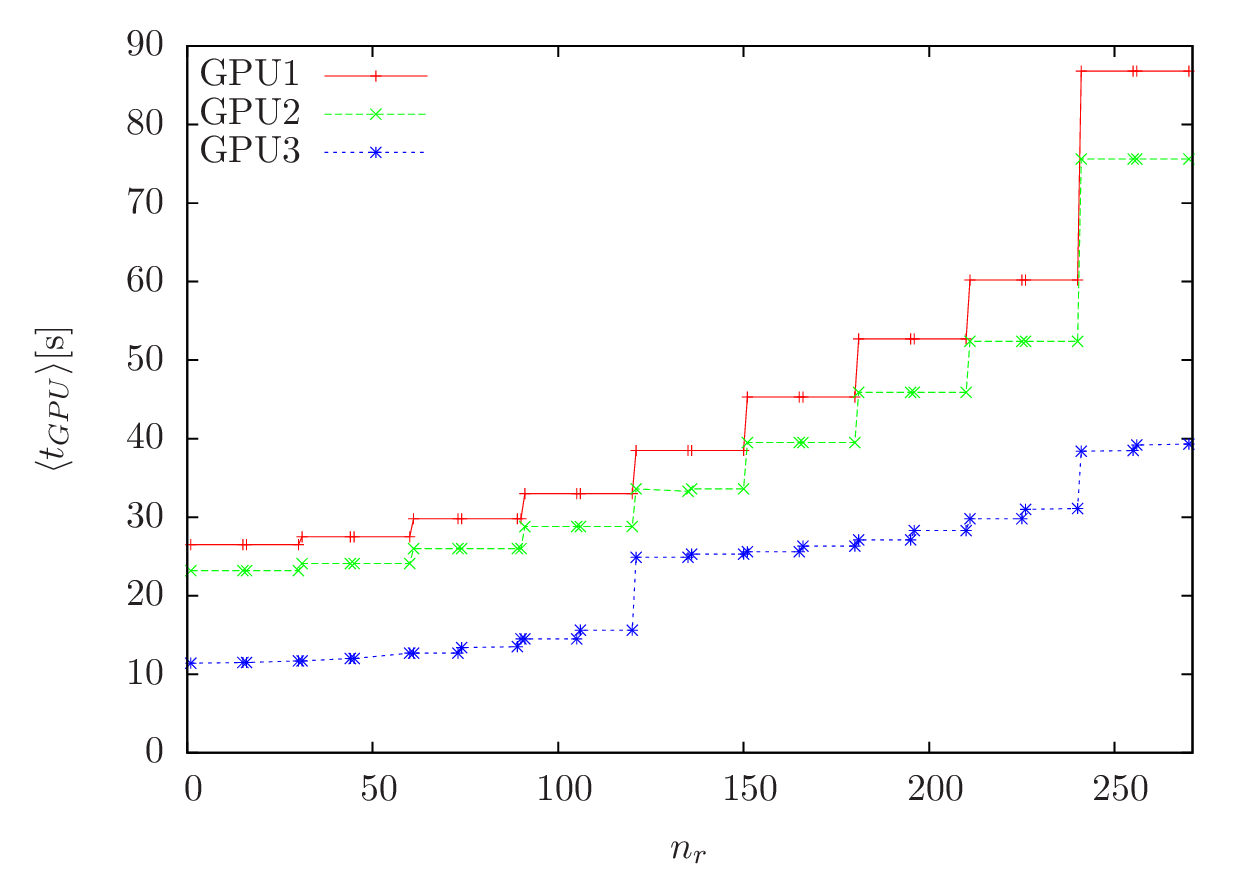}
  \caption{Average GPU time $\langle t_{GPU}\rangle$ in seconds vs. number of replicas $n_r$ for the improved GPU version.}
  \label{fig:runtimes2}
\end{figure}
The steplike shape of the curve in Fig.~\ref{fig:runtimes2} every 30 replica for GPU1 and GPU2 coincides with the number of SMs, meaning that with every additional thread block the overall speed-up drops until each of the 30 multiprocessors again is equally busy. 
Also noticeable is that for GPU1 and GPU2 the first maximum of the speed-up factor is reached for 120 thread blocks. 
That means all cores are equally busy, but apparently there seems to be plenty of latency in memory operations.
Thus, the overall speed-up is not affected by adding the same amount of work to each multiprocessor, up to 240 thread blocks in total. 
For GPU3 the increase in kernel runtime occurs every 15 thread blocks, corresponding to the number of SMs in the given card. 
Consequently, to maximize the benefit from GPU implementations it is necessary to keep all multiprocessors on the GPU equally busy.

\begin{table}[t]
  \centering
  \caption{Overview of maximum achieved speed-ups -- $\max\left(S_p(n_r)\right)$ -- for the two different GPU implementations, compared to the single-core CPU implementation.}
  \begin{tabular}{l r r}
  & naive & improved \\
  \hline \hline \\
  GPU1 & $6.1\times$ & $68\times$ \\
  GPU2 & $7.2\times$ & $78\times$ \\
  GPU3 & $9\times$  & $130\times$ \\
  \hline
  \end{tabular}
  \label{table:speedups}
\end{table}

Table \ref{table:speedups} is a summary of the maximum speed-up factors achieved in our simulations, employing the two different implementations.
The ratio of speed-ups from GPU1 and GPU2 is nearly the same as the ratio of their clock speeds.
Whereas GPU3 with a similar clock speed shows significant speed-ups, which originate from the difference in the chip design of the two GPU generations.

\section{Summary}
\label{sec:summary}
In this paper, we have shown that replica-exchange Monte Carlo simulations of off-lattice polymer models can be performed quite efficiently on GPUs.
Even for off-lattice polymer models this is a suitable approach.
Already with a very simple naive porting of CPU code to the GPU, we find considerable performance gains of factors about 6--9 compared to a single CPU implementation.
Utilizing the unique architecture of GPUs, with its different memory layers and the ability to schedule a massive amount of threads, we improved the GPU program to attain speed-up factors of around 70 for mainstream GPUs like the GTX285, and even factors up to 130 with NVIDIA's new generation Fermi-based GTX480 card.
It should be noted that our implementations represent a rather basic level of utilizing the advantages of GPU architectures.

Furthermore it is possible to access multiple cards in a single workstation from one and the same program with no extra effort.
Also nodes of established cluster computers can be equipped with GPUs, a combination of the traditional message  passing interface (MPI) and CUDA is used in such a scenario.
Thus GPUs promise great gains in productivity and might help building the next-generation supercomputers.

\section*{Acknowledgements}

We acknowledge support by the German-Israeli program ``Umbrella'' under Grant No.\ HPC\_2, the German Research Foundation (DFG) under Grant Nos.\ JA 483/24-2/3, the Leipzig Graduate School of Excellence ``BuildMoNa'', the German-French DFH-UFA PhD College under Grant No. CDFA-02-07, and by the Forschungszentrum J\"ulich for supercomputing time grants jiff39 and jiff43.


\end{document}